\documentclass[epj]{webofc}
\usepackage[varg]{txfonts}   
\woctitle{MESON2014 - the 13$^\textrm{th}$ International Workshop on Meson Production, Properties and Interaction}
\begin{document}
\selectlanguage{english}
\title{Dalitz plot analysis of \mbox{\boldmath $D^0 \to K^0_S \pi^+ \pi^-$} decays in a factorization \\ approach}

\author{Leonard Le\'sniak\inst{1}\fnsep\thanks{\email{Leonard.Lesniak@ifj.edu.pl}} \and
        Jean-Pierre Dedonder\inst{2} \and
        Robert Kami\'nski\inst{1} \and
        Benoit Loiseau \inst{2}
}

\institute{Division of Theoretical Physics, The Henryk Niewodnicza\'nski Institute of Nuclear Physics, Polish Academy of Sciences, 31-342 Krak\'ow, Poland 
\and
           Sorbonne Universit\'es, Universit\'e Pierre et Marie Curie, Sorbonne Paris Cit\'e, Universit\'e Paris Diderot, et IN2P3-CNRS, UMR 7585, Laboratoire de Physique Nucl\'eaire et de Hautes \'Energies,  4 place Jussieu,  ~~~~~~~       75252 Paris, France 
          }

\abstract{A quasi two-body QCD factorization is used to study the $D^0 \to K^0_S \pi^+ \pi^-$ decays.
The presently available 
high-statistics Dalitz plot data of this process measured by the Belle and 
BABAR Collaborations are analyzed together with the $\tau^- \to K^0_S \pi^-\nu_\tau$ 
decay data. The 
total experimental branching fraction is also included in the fits which show a very good 
overall agreement with the experimental Dalitz plot density distributions. The branching fractions of the 
dominant channels compare well with those of the isobar Belle or BABAR models. We show that 
the branching fractions corresponding to the annihilation amplitudes are 
significant.} 
\maketitle
\section{Introduction}
\label{intro}

Preliminary results of the Dalitz plot analysis of the $D^0 \to K^0_S \pi^+ \pi^-$ decays have been
already shown during the MESON2012 Workshop \cite{Meson2012}. One of our aims was 
the construction 
of D-decay amplitudes in which unitarity is preserved in two-body subchannels like in $K^0 \pi$
$S$-wave and $\pi\pi$ $S$- and $P$-waves. Below we present some new results 
recently published in Ref. \cite{DKLL}. 
Starting from the weak effective Hamiltonian, 28 tree and annihilation (W-exchange) amplitudes
 build up the full $D^0 \to K^0_S \pi^+ \pi^-$ amplitude. The meson-meson final state 
interactions
 are described by the kaon-pion and pion scalar and vector form factors for the $S$ and $P$ waves
 and by Breit-Wigner formulae for the $D$ waves. Unitarity, analyticity and chiral symmetry are
 used to constrain functional forms of the form factors which group several resonances in
 a given
 partial wave. This, together with charge symmetry, allows to reduce the 27 non-zero amplitudes
 into 10 effective amplitudes depending on 33 free parameters.

\section{Meson-meson form factors and effective mass distributions}
\label{sec-1}
\begin{figure}[h] 
\begin{center}
\includegraphics[scale = 0.3]{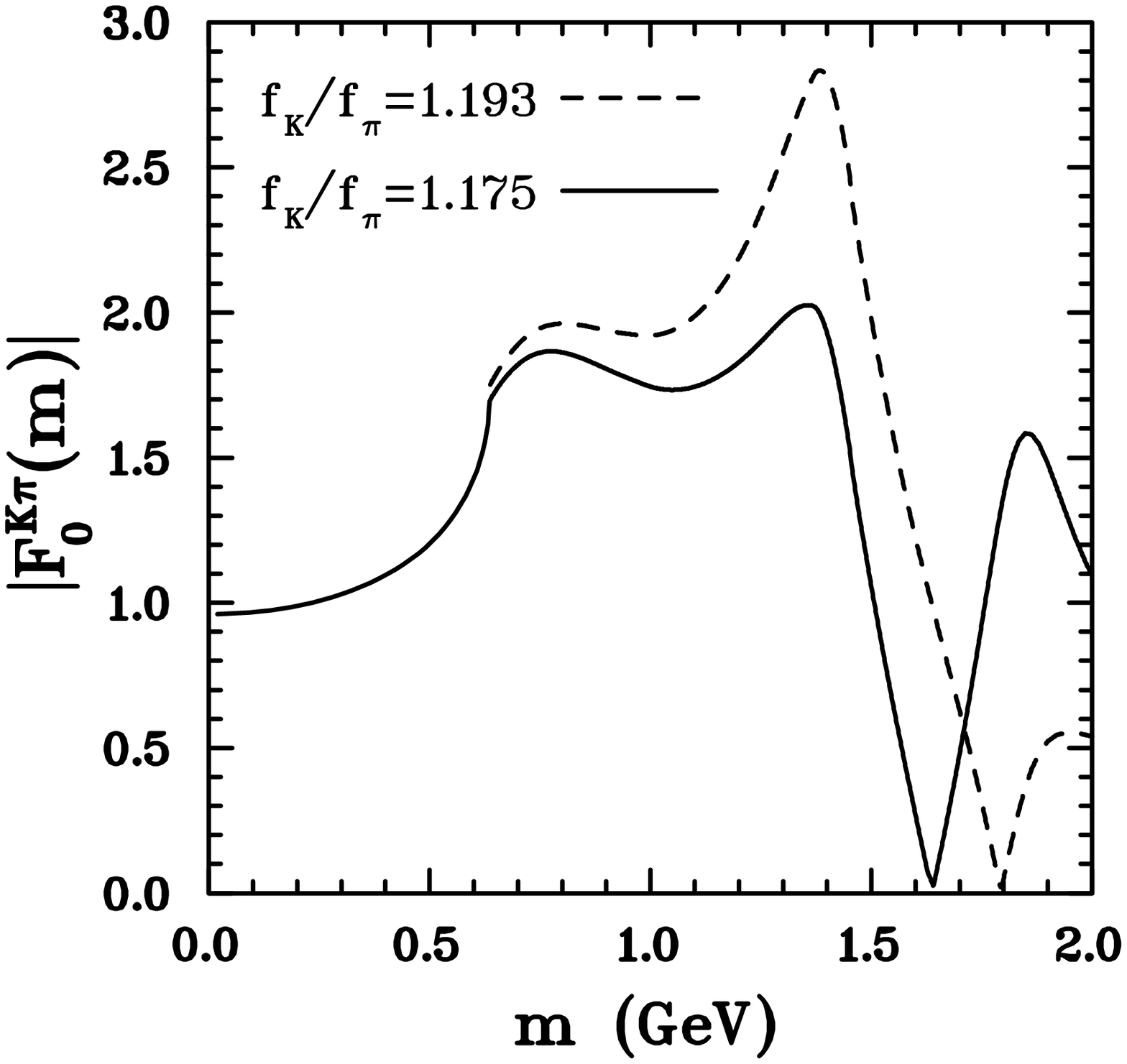}~~~~
\includegraphics[scale = 0.3]{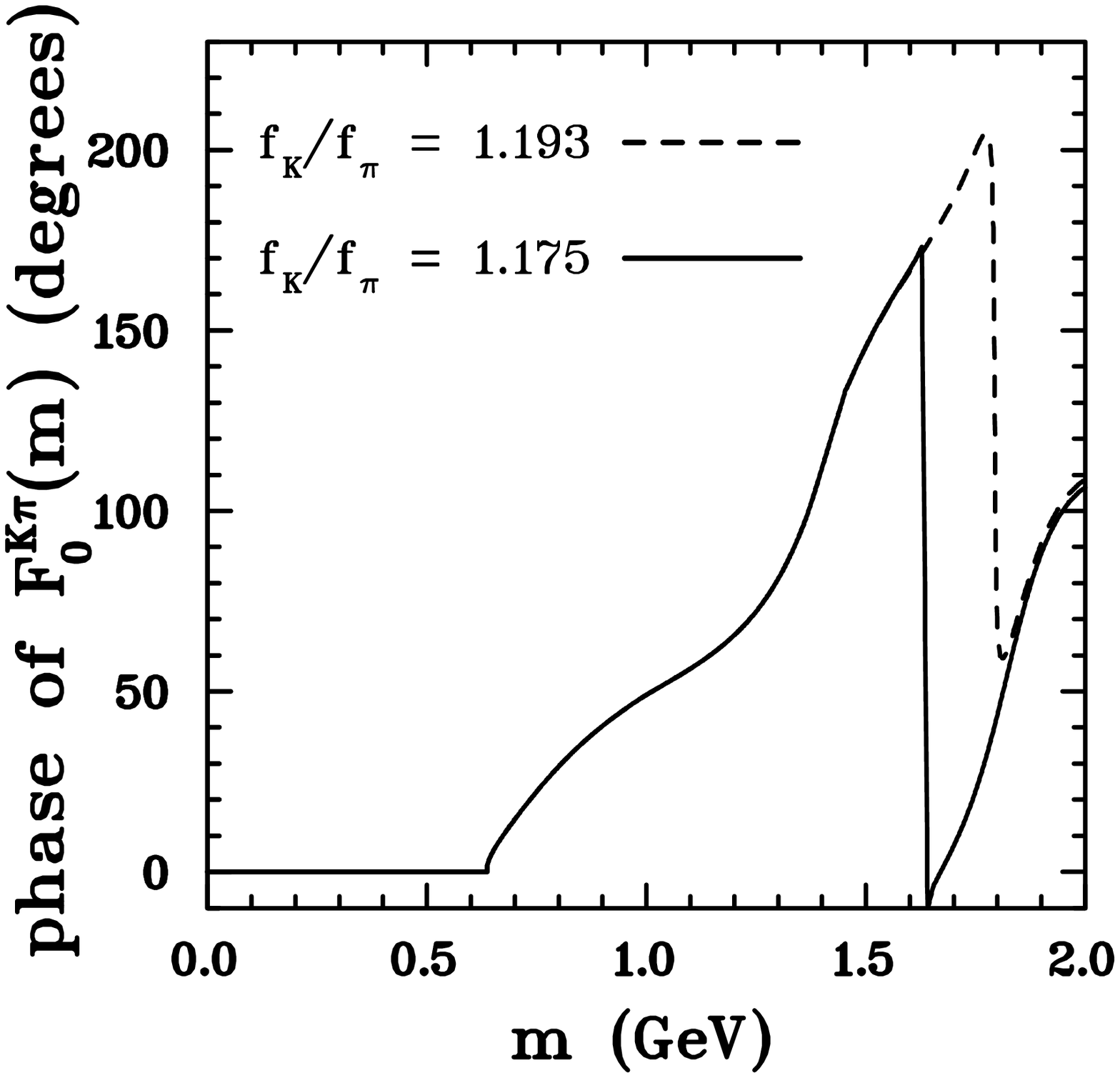}~~~~

\caption{The modulus (left panel) and the phase (right panel) of the 
 $K \pi$ scalar form factor  $F^{K\pi}_0$ as function of the
$K\pi$ effective mass for two values of the $f_K/f_{\pi}$ ratio, where $f_K$ and $f_{\pi}$ are the
kaon and pion decay constants.
}
\label{F0Kpi}
\end{center}
\end{figure}

\begin{figure}[h] 
\begin{center}
\includegraphics[scale = 0.3]{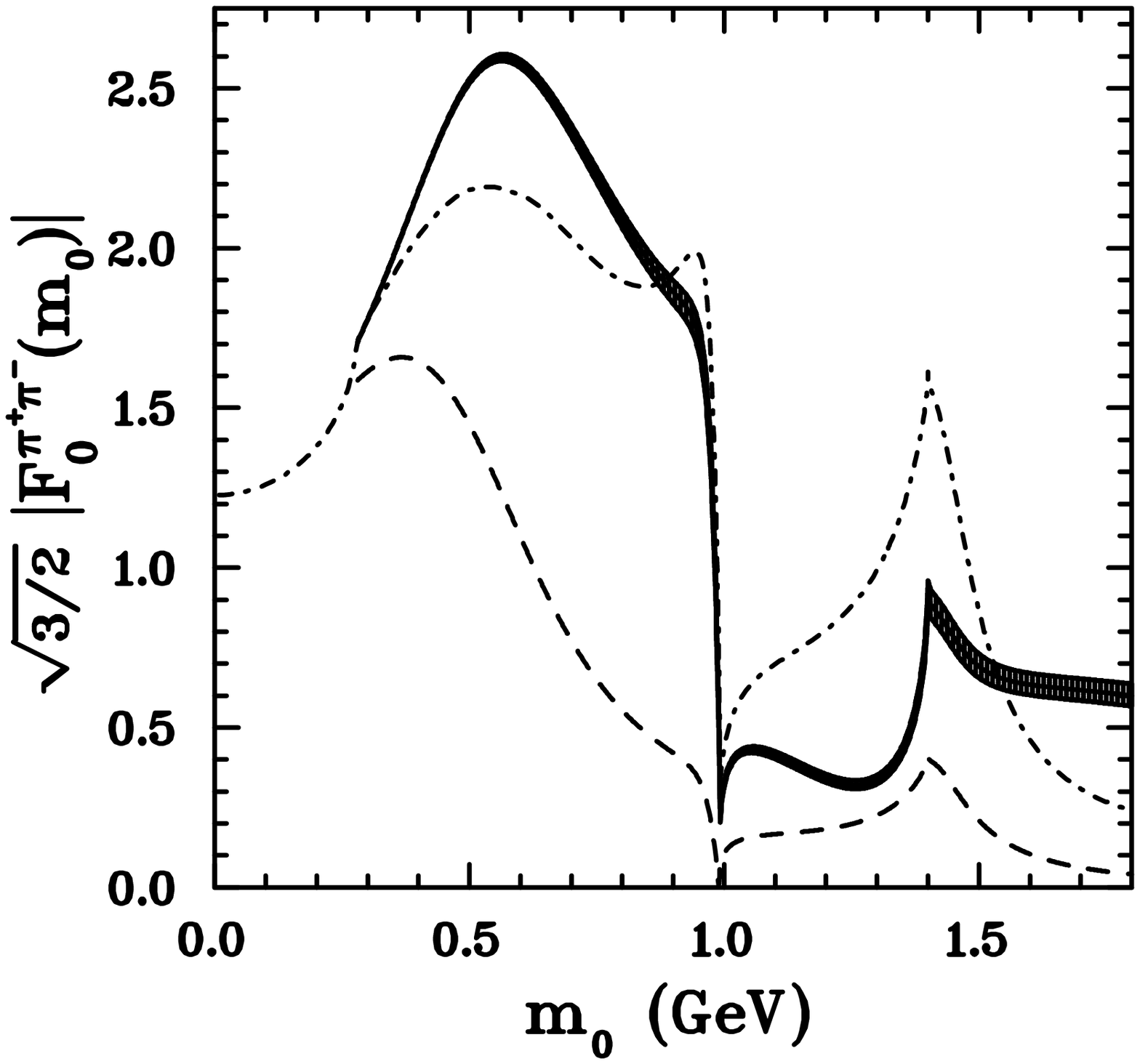}~~~~
\includegraphics[scale = 0.3]{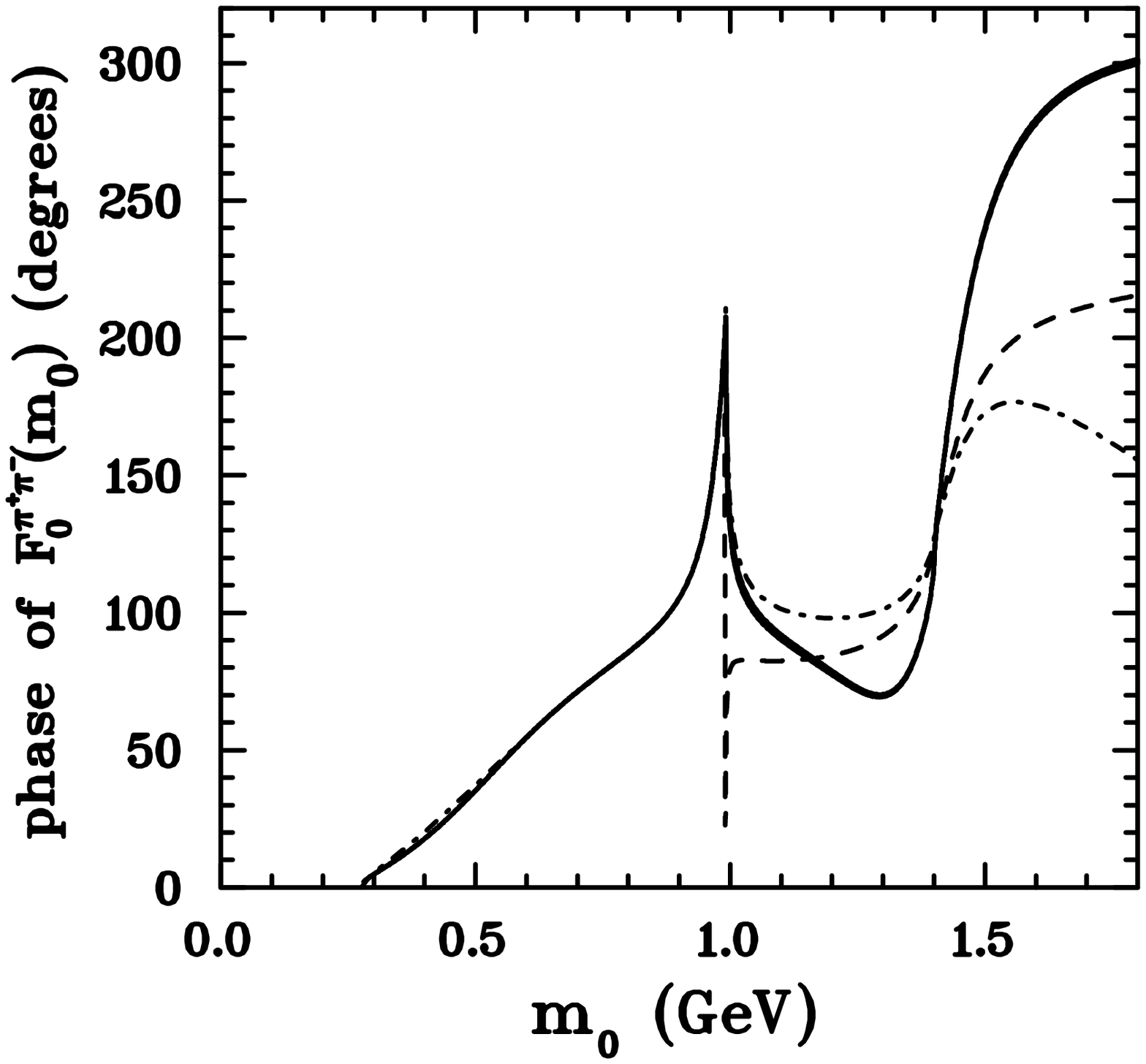}

\caption{The modulus (left panel) and phase (right panel) of the pion scalar
form factor$F^{\pi^+ \pi^-}_0(m_0)$, resulting from the fit to the Belle
data, are plotted as functions of the $\pi \pi$ effective mass.
The dark bands represent their variations when the parameters vary
within their errors.
They are compared with the same form factor obtained in Ref.~\cite{DedonderPol} with
different parameters (dashed line) and with the form factor calculated
from the Muskhelishvili-Omnès equations ~\cite{Moussallam_2000} (dotted-dashed line).
}
\label{F0pipi}
\end{center}
\end{figure}

The final state strong interactions between mesons influence the functional dependence of the 
meson-meson form factors on the effective mass variables. An important role in the decay amplitude
is played by the scalar form factors. In 
Fig.~\ref{F0Kpi} the $K\pi$ scalar form factor is shown. On the left panel one can see two
maxima lying below 1.5 GeV which correspond to the strange scalar resonances $K^*(800)$ and $K^*(1430)$. In Fig.~\ref{F0pipi}
the pion scalar form factor is plotted and three scalar resonances are jointly described. In the left
panel the two maxima correspond to the $f_0(500)$ and $f_0(1400)$ resonances while the deep minimum
at about 1 GeV is related to $f_0(980)$.

A joint $\chi^2$ fit to the Belle Dalitz plot density distribution 
\cite{A.Poluektov_PRD81_112002_Belle}, to the $\tau^- \to K^0_S \pi^- \nu_\tau$ decay data
\cite{EpifanovPLB654} and to the total experimental branching fraction is carried out to fix the 33 
free parameters.
These are mainly related to the strengths of the scalar form factors and to unknown meson to meson 
transition form factors at a large momentum transfer squared equal to the $D^0$ mass squared.
The present model Dalitz plot density distribution is shown in the left panel of Fig. 3.
The Dalitz variables projections seen in the right panel of Fig. 3 and in Fig. 4 indicate that
 a good overall agreement to the Belle Dalitz plot density distribution is achieved.
The scalar and vector $K\pi$ form factors have been constrained by a fit to the
$\tau^- \to K^0_S \pi^- \nu_\tau$ data. The red solid line in Fig. 5 shows the model 
distribution compared to the data of Ref.~\cite{EpifanovPLB654}. 
A separate good quality fit to the BABAR isobar model distributions \cite{SanchezPRL105_081803} has been
performed leading to a similar set of parameters as that obtained for the fit to the Belle data.  

The four most important channel contributions to the branching fraction are given in Table 1.
The $S$-wave channel branching fractions are sizable.
The last column gives the lowest values of the annihilation parts. These values are significant
in comparison with the tree branching fractions and indicate the importance of the W-exchange 
contributions to the full decay amplitude.

Our $D^0 \to K^0_S \pi^+ \pi^-$ decay amplitude could be a useful input for determinations of
the $D^0-\overline{D}^0$ mixing parameters and of the Cabibbo-Kobayashi-Maskawa angle $\gamma$ 
(or $\phi_3$). Upon request, we can provide numerical values for our amplitudes.
\begin{table}
\centering
\caption{Branching fractions (Br) for different quasi two-body channels calculated for the best 
fit to the Belle data~\cite{A.Poluektov_PRD81_112002_Belle}.
The branching fractions for the tree amplitudes (tree) and the lowest values for the 
annihilation amplitudes (Ann. low.) are also given. The errors are statistical. All numbers are in per cent.}
\label{tab-1}       
\begin{tabular}{llll}
\hline
Channel & Br~(\%) & Br~(tree) & Ann. low.  \\\hline
$[K^0_S\pi^-]_S\pi^+$ & 25.0$\pm$3.6 & $8.2\pm0.1$ & 7.9$\pm$0.1 \\
$K^0_S[\pi^-\pi^+]_S$ & 16.9$\pm$1.3 & $14.7\pm0.2$ & 2.9$\pm$0.1 \\
$[K^0_S\pi^-]_P\pi^+$ & 62.7$\pm$4.5 & $24.7\pm5.7$ & 8.7$\pm$3.0 \\
$K^0_S[\pi^-\pi^+]_P$ & 22.0$\pm$1.6 & $4.4\pm0.1$ & 6.7$\pm$0.04 \\\hline
\end{tabular}

\end{table}
\begin{figure}
\begin{center}
  \includegraphics[height=.3\textheight]{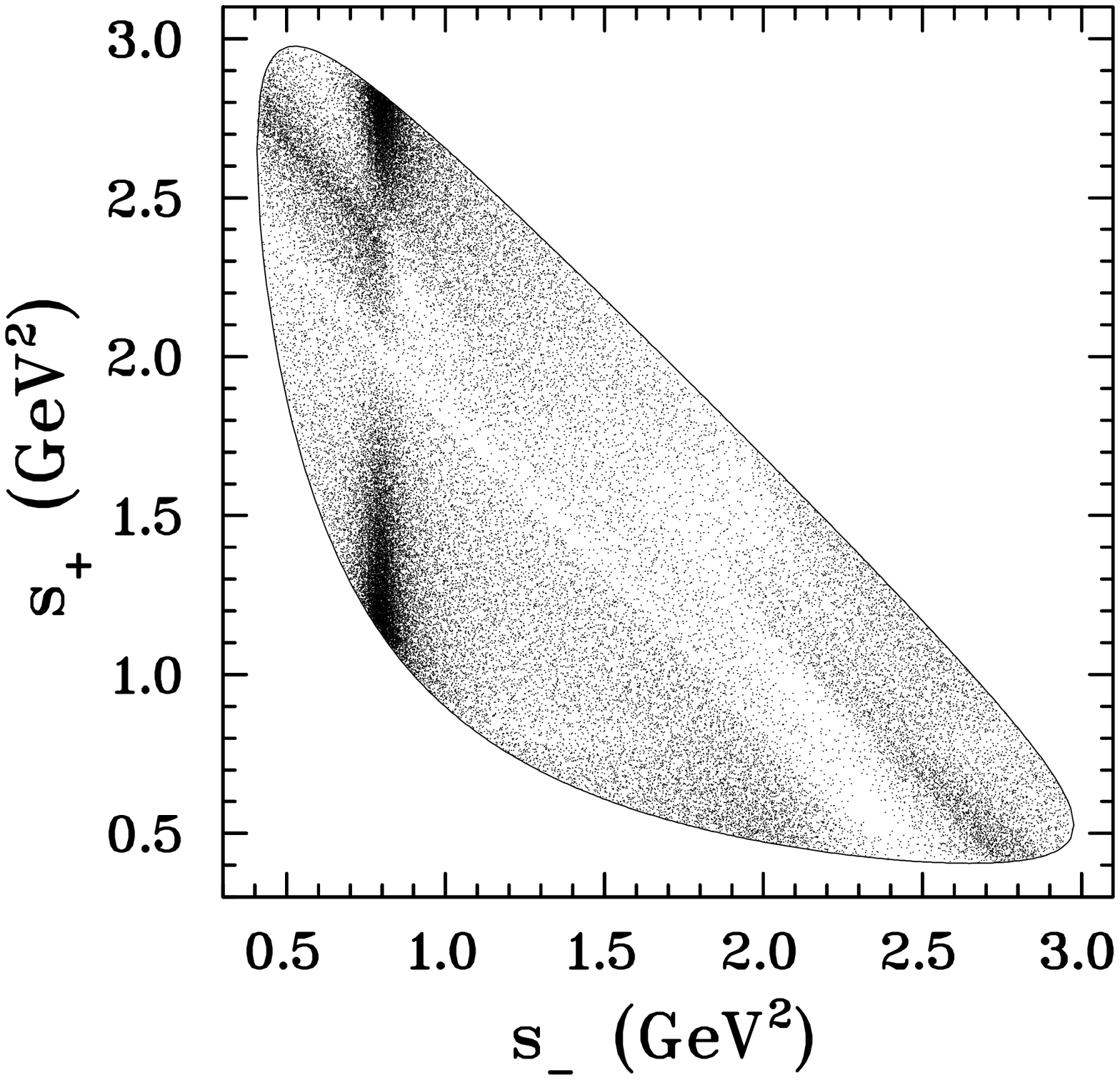}~~~~
   \includegraphics[height=.3\textheight]{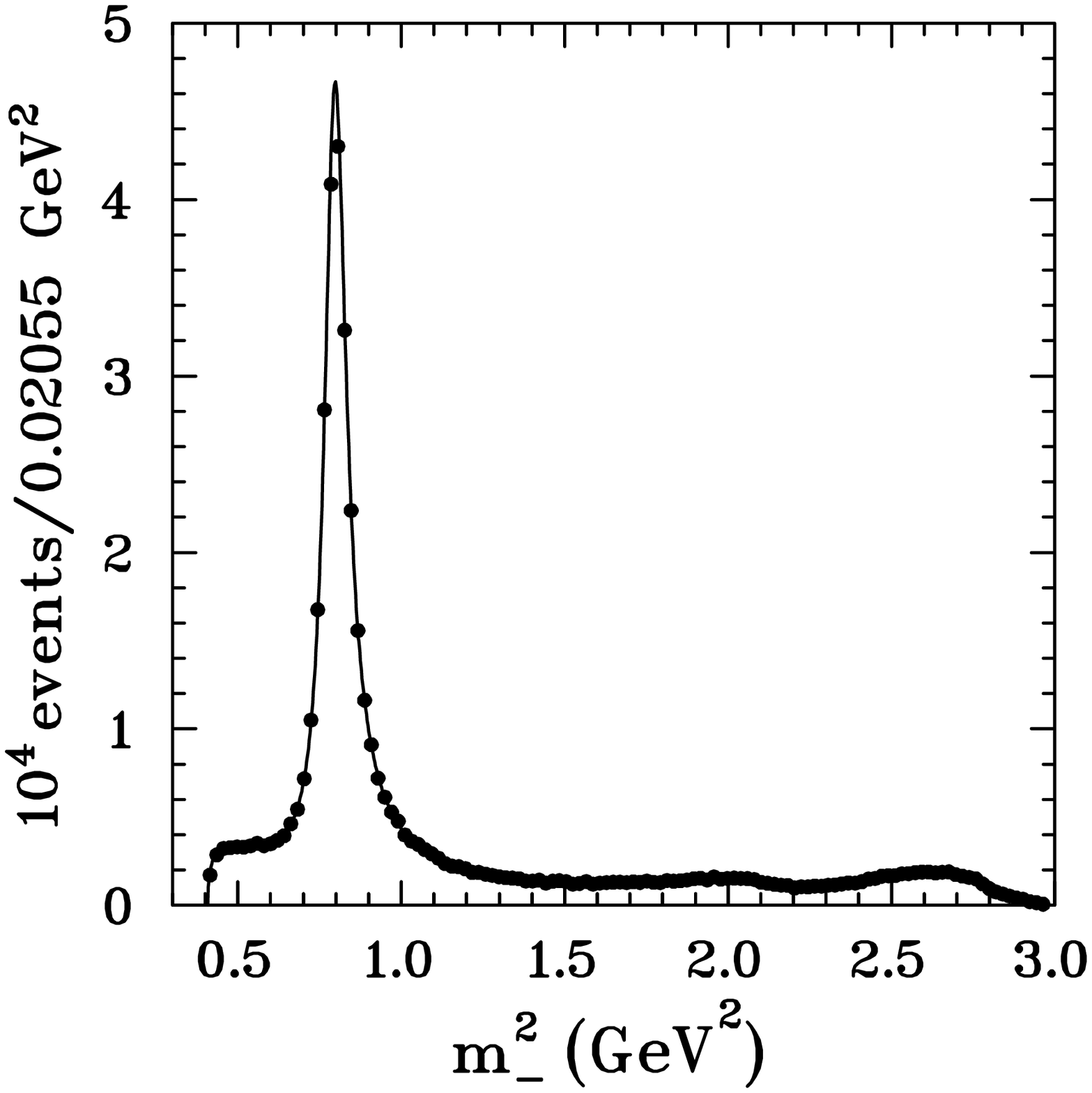}~~
\caption{Left panel: Dalitz plot distribution from the fit to the Belle 
data~\cite{A.Poluektov_PRD81_112002_Belle}. Comparison of the $K^0_S\pi^-$ effective mass squared distributions
for our model (solid curve) with the Belle data~\cite{A.Poluektov_PRD81_112002_Belle} (points with error 
bars). 
}\label{Polmmin}
\end{center}
\end{figure}

\begin{figure}
\begin{center}
  \includegraphics[height=.28\textheight]{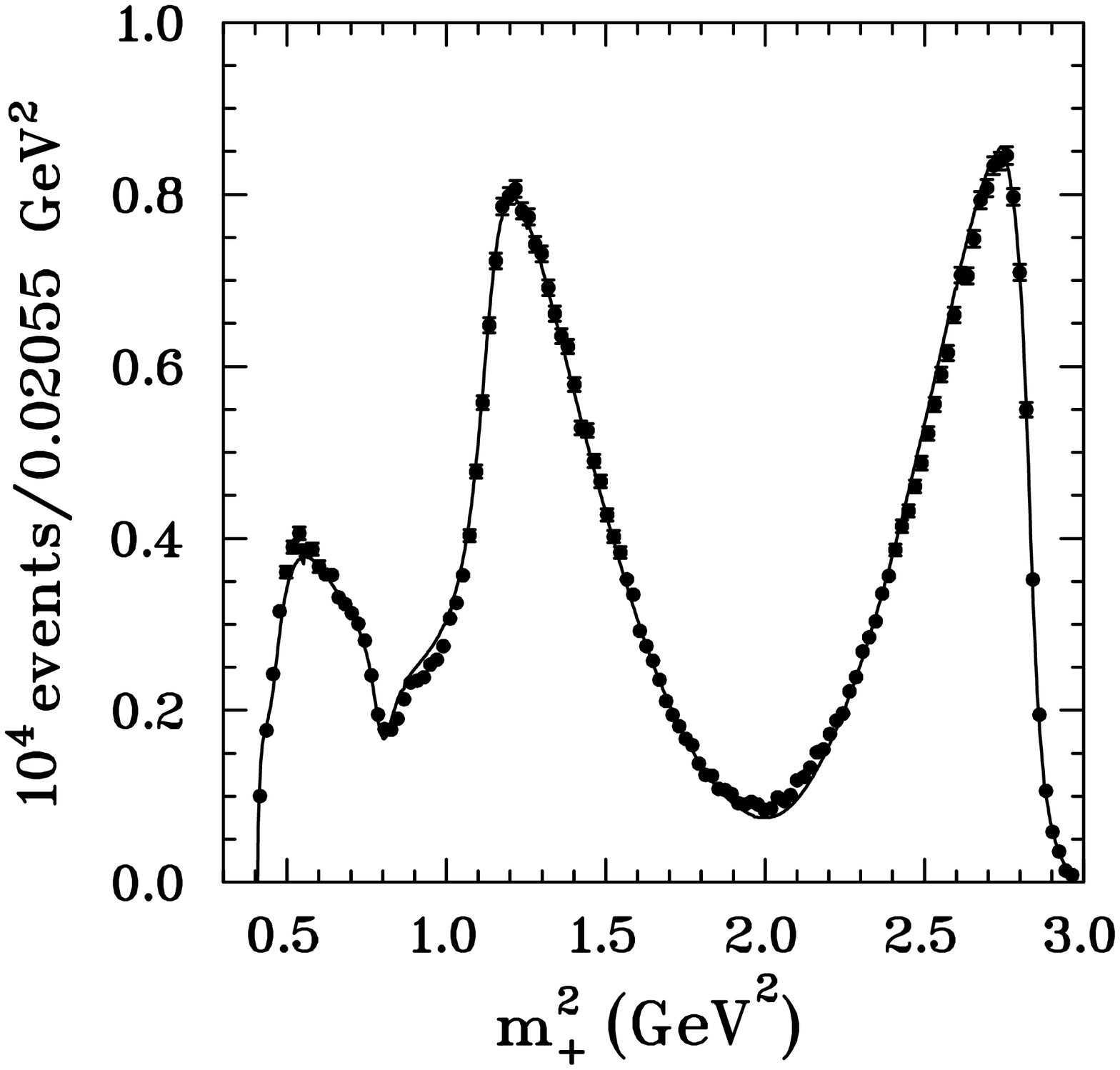}~~~~
   \includegraphics[height=.28\textheight]{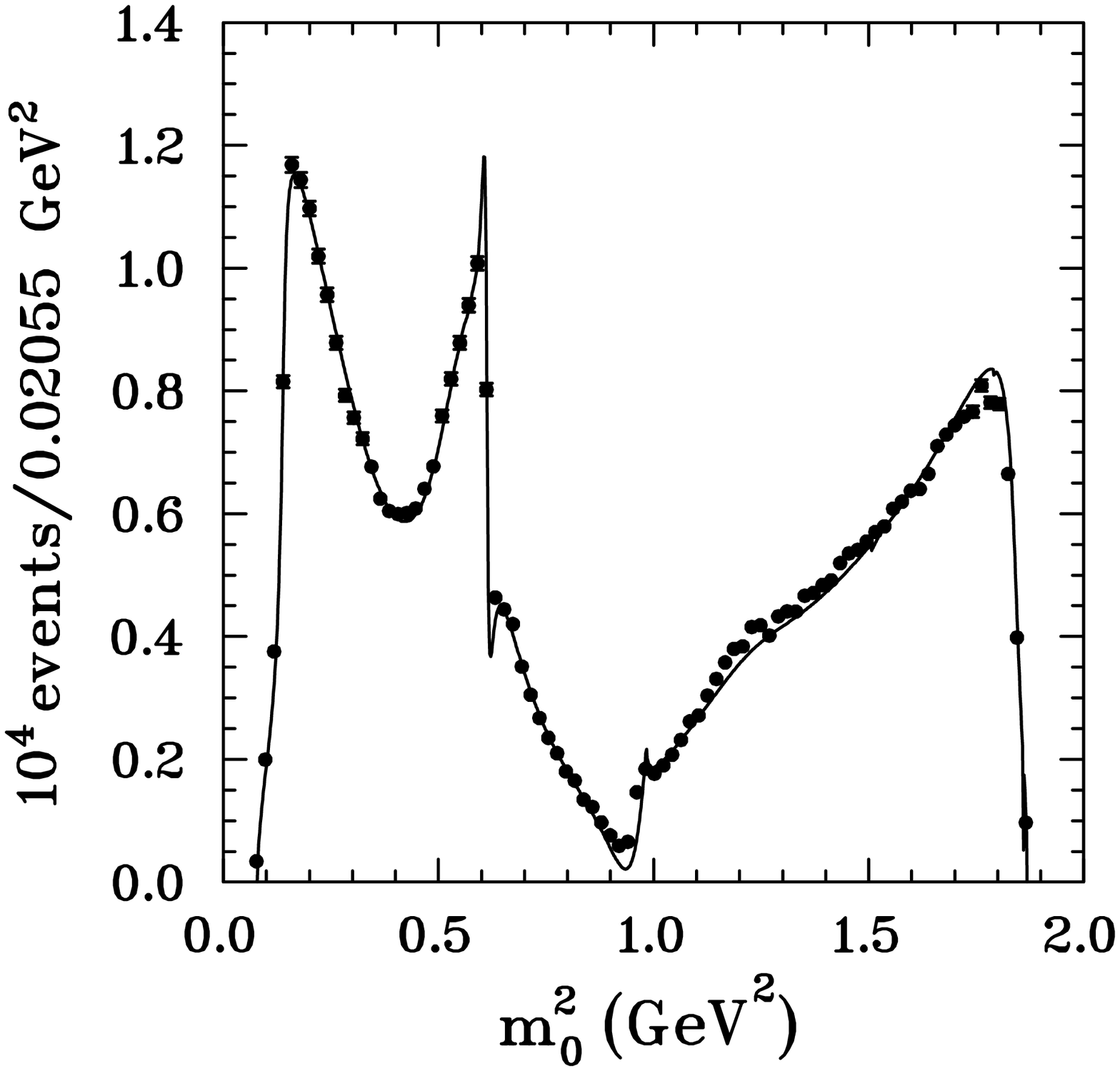}~~
\caption{Left panel: comparison of the $K^0_S\pi^+$ effective mass squared distributions
for the best fit (solid curve) with the Belle data~\cite{A.Poluektov_PRD81_112002_Belle}. 
Right panel: as in left panel but for the $\pi^+\pi^-$ effective mass squared.} 
\label{Polmplus}
\end{center}
\end{figure}

\begin{figure} 
\begin{center}
\sidecaption

  \includegraphics[height=.3\textheight]{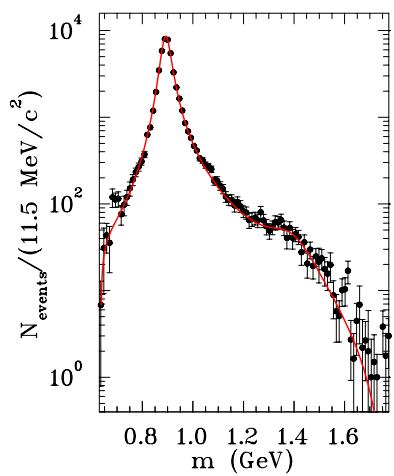}

\caption{Comparison of the fit of the model (red solid line) with the $K^0_S \pi^-$ effective mass
distribution of the Belle data on the $\tau^- \to K^0_S \pi^-\nu_\tau$ decays~\cite{EpifanovPLB654}}
\label{tau}
\end{center}
\end{figure}

\begin{acknowledgement}
This work has been partially supported by a grant from the French-Polish 
exchange program COPIN/CNRS-IN2P3, collaboration 08-127.
\end{acknowledgement}

\end{document}